\documentclass[conference]{IEEEtran}

\usepackage{booktabs}
\usepackage{tikz}

\begin{document}

\title{An Overview of the Threat of Voice Cheater}
\title{\huge An Overview of Vulnerabilities of Voice Controlled Systems}

\author{\IEEEauthorblockN{Yuan Gong}
\IEEEauthorblockA{Computer Science and Engineering\\
University of Notre Dame, IN 46556\\
Email: ygong1@nd.edu}
\and
\IEEEauthorblockN{Christian Poellabauer}
\IEEEauthorblockA{Computer Science and Engineering\\
University of Notre Dame, IN 46556\\
Email: cpoellab@nd.edu}
}

\maketitle

\begin{abstract}

Over the last few years, a rapidly increasing number of Internet-of-Things (IoT) systems that adopt voice as the primary user input have emerged. These systems have been shown to be vulnerable to various types of voice spoofing attacks. However, how exactly these techniques differ or relate to each other has not been extensively studied. In this paper, we provide a survey of recent attack and defense techniques for voice controlled systems and propose a classification of these techniques. We also discuss the need for a universal defense strategy that protects a system from various types of attacks.

\end{abstract}


\IEEEpeerreviewmaketitle

\section{Introduction}
\begin{figure}[t]
  \centering
  \includegraphics[width=8cm]{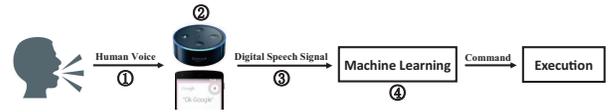}
  \caption{An illustration of a typical voice controlled system. The device captures the human voice, converts it into a digital speech signal, and feeds it into a machine learning model. The corresponding command is then executed by the connected IoT devices. Potential points of attack in this scenario include: \textcircled{1}: spoofing the system using previously recorded audio, \textcircled{2}: hacking into the operating system to force the voice-driven software to accept commands erroneously, \textcircled{3}: emitting carefully designed illegitimate analog signals that will be converted into legitimate digital speech signals by the hardware, and \textcircled{4}: using carefully crafted speech adversarial examples to fool the machine learning model.}
  \label{fig:system}
\end{figure}

An increasing number of IoT systems rely on voice input as the primary user-machine interface. For example, voice-controlled devices such as Amazon Echo, Google Home, Apple HomePod, and Xiaomi AI allow users to control their smart home appliances, adjust thermostats, activate home security systems, purchase items online, initiate phone calls, and complete many other tasks with ease. In addition, most smartphones are also equipped with smart voice assistants such as Siri, Google Now, and Cortana, which provide a convenient and natural user interface to control smartphone functionality or IoT devices.
Voice-driven user interfaces allow hands-free and eyes-free operation where users can interact with a system while focusing their attention elsewhere.
A block diagram of a typical voice controlled system (VCS) is shown in Figure~\ref{fig:system}.

Despite their convenience, VCSs also raise new security concerns. One such concern is their vulnerability to a voice replay attack~\cite{14chen2017you}, i.e., an attacker can replay a previously recorded voice to make the IoT system perform a specific malicious action. Such malicious actions include the opening and unlocking of doors, making unauthorized purchases, controlling sensitive home appliances (e.g., security cameras and thermostats), and transmitting sensitive information. While a simple voice replay attack is relatively easy to detect by a user, and therefore presents only a limited threat, recent studies have pointed out more concerning and effective types of attacks, including self-triggered attacks~\cite{08diao2014your,21jang2014a11y}, inaudible attacks~\cite{01zhang2017dolphinattack,11kasmi2015iemi}, and human-imperceptible attacks~\cite{03carlini2016hidden,05gong2017crafting,12carlini2018audio}. These attacks are very different from each other in terms of their implementation, which requires different domain knowledge in areas such as the operating system, signal processing, and machine learning. However, how exactly they differ or relate to each other has not been extensively studied. As a consequence, to the best of our knowledge, current defense techniques only aim to defend attacks of one specific category, with the assumption that the defender knows the details of the attacking technology. From a security standpoint, this is deeply unsatisfactory. Therefore, in this paper, we provide a survey of recent attack and defense techniques for voice controlled systems, discuss their relationships, and propose a classification of these techniques. We further discuss a potential universal defense strategy for different types of attacks. We expect that the analysis and discussion in this paper will provide useful insights for future studies and efforts in building secure voice-driven IoT systems.

\section{Voice-Based Attacks}
With the rapidly growing popularity and functionality of voice-driven IoT devices, the potential of voice-based attacks becomes a non-negligible security risk. As discussed in~\cite{04alepis2017monkey,08diao2014your,07lei2017insecurity}, an attack may lead to severe losses, e.g., a burglar could enter a house by tricking a voice-based smart lock or an attacker could make unauthorized purchases and credit card charges using a voice-based system. Such attacks can be very simple and often difficult or even impossible to detect by humans and voice attacks can be hidden by other sounds or embedded into audio and video recordings. Further, it is also very easy to scale up such attacks, e.g.,  
a hidden malicious audio sample in a YouTube video could simultaneously target millions of devices. 

Although the implementations of existing attack techniques may be very different, their goals are the same: generating a signal that leads a voice controlled system to execute a specific malicious command that the user cannot detect or recognize. In the following sections, we first introduce representative state-of-the-art attack approaches according to the type of implementation. We then further discuss the positives and negatives of each approach and how they relate to each other. The attacker performance discussed in this section is evaluated and reported by the original publication. Due to rapid changes of cloud-based systems, the attacker performance is also likely to change over time.

\subsection{Attack Classification Based On Implementation}
\begin{table*}[t]
\centering
\caption{Representative voice attack techniques}
\label{tab:attack}
\begin{tabular}{@{}p{3.8cm}p{2.1cm}p{2.3cm}p{8.0cm}@{}}
\toprule
\multicolumn{1}{c}{Attack Name} & \multicolumn{1}{c}{Attack Type} & \multicolumn{1}{c}{Adversary's Knowledge} & \multicolumn{1}{c}{Implementation} \\ \midrule

GVS Attack~\cite{08diao2014your} & Operating System & White box 
& Continuously analyze the environment and conduct voice replay attack using built-in microphone when opportunities arise.\\

A11y Attack~\cite{21jang2014a11y} & Operating System & White box 
& Collect the voice of a user and conduct self-replay attack as a background service.\\

Monkey Attack~\cite{04alepis2017monkey} & Operating System & White box 
& Bypass authority management of the OS and conduct interactive voice replay attack to execute more advanced commands.\\

Dolphin Attack~\cite{01zhang2017dolphinattack} & Hardware & White box 
& Emit ultrasound signal that can be converted into a legitimate speech digital signal by the MEMS microphone.\\

IEMI Attack~\cite{11kasmi2015iemi} & Hardware & White box 
& Emit AM-modulated signal that can be converted into a legitimate speech digital signal by the wired microphone-capable headphone.\\

Cocaine Noodles~\cite{02vaidya2015cocaine} & Machine Learning & Black box & Similar to the hidden voice command. \\

Hidden Voice Command~\cite{03carlini2016hidden} & Machine Learning & Black \& White box 
& Mangle malicious voice commands so that it retains enough acoustic features for the ASR system, but becomes unintelligible to humans.\\

Houdini~\cite{06cisse2017houdini} & Machine Learning & Black \& White box 
& Produce sound that is almost no different to normal speech, but fails to be recognized by both known or unknown ASR systems.\\

Speech Adversarial Example~\cite{05gong2017crafting} & Machine Learning & White box 
& Produce sound that is over 98\% similar to any given speech, but makes the DNN model fail to recognize the gender, identity, and emotion. \\

Targeted Speech Adversarial    Example~\cite{12carlini2018audio} & Machine Learning & White box 
& Produce sound that is over 99.9\% similar to any given speech, but transcribes as any desired malicious command by the ASR. \\ 
\bottomrule
\end{tabular}
\end{table*}

\subsubsection{Basic Voice Replay Attack}

It is widely known that voice controlled systems are vulnerable to voice replay attacks, i.e., an attacker can replay a previously recorded voice to make a system perform a specific action~\cite{14chen2017you,09petracca2015audroid}, e.g., as demonstrated previously with the popular Amazon Alexa technology~\cite{07lei2017insecurity}. A shortcoming of the basic voice replay attack is that it is easy to detect and therefore has a limited practical impact. Nevertheless, as shown later in this section, voice replay attacks are the basis of other more advanced and dangerous attacks.

\subsubsection{Operating System Level Attack}

Compared to basic voice replay attacks, an operating system (OS) level attack 
exploits vulnerabilities of the OS to make the attack self-triggered and more imperceptible. Representative examples of this are the A11y attack~\cite{21jang2014a11y}, GVS-Attack~\cite{08diao2014your}, and the approach presented in~\cite{04alepis2017monkey}. In~\cite{21jang2014a11y}, the authors propose a malware that collects a user's voice and then performs a self-replay attack as a background service. In~\cite{08diao2014your}, the authors further verify that the built-in microphone and speaker can be used simultaneously and that the use of the speaker does not require user permission on Android devices. They take advantage of this and propose a zero-permission malware, which continuously analyzes the environment and conducts the attack once it finds that no user is nearby. The attack uses the device's built-in speaker to replay a recorded or synthetic speech, which is then accepted as a legitimate command. This self-triggered attack is thus more dangerous and practical. While this attack can still be detected by the user, the authors point out that if the malware has high permissions, it is even possible for it to import an audio file to the microphone without playing it, which can make the attack completely inaudible. In~\cite{04alepis2017monkey}, the authors analyze the permission vulnerability to the voice attack in detail and propose an approach to bypass the permission management of the Android system. The authors also find that some malicious actions require a multiple-step command and further propose an interactive attack that can execute more advanced commands.

\subsubsection{Hardware Level Attack} 

A hardware level attack replays a synthetic non-speech analog signal instead of human voice. 
The analog signal is carefully designed according to the characteristics of the hardware (e.g., the analog-digital converter). The signal is inaudible, but can be converted into a legitimate digital speech signal by the hardware. Representative approaches are the Dolphin attack~\cite{01zhang2017dolphinattack} and the IEMI attack~\cite{11kasmi2015iemi}. In~\cite{01zhang2017dolphinattack}, the authors utilize the non-linearity of a Micro Electro Mechanical Systems (MEMS) microphone over ultrasounds and successfully generate inaudible ultrasound signals that can be accepted as legitimate target commands. Generating such ultrasound signals requires a special device that includes a controller (e.g., another smartphone), an amplifier, and an ultrasonic transducer. The longest attack distance is 175cm. In~\cite{11kasmi2015iemi}, the authors take advantage of the fact that a wired microphone-capable headphone can be used as a microphone and an FM antenna simultaneously and demonstrate that it is possible to trigger voice commands remotely by emitting a carefully designed inaudible AM-modulated signal. This attack is only effective when the wired headphone is plugged into the device. A limitation of hardware level attacks is that generating the attack signal requires special devices, and also some preconditions must be met (e.g., the victim device needs to be in the attack range and the microphone needs to be plugged in). While the synthetic signal is inaudible to the user, the signal generator might still be noticed by the user.

\begin{figure}[h]
  \centering
  \includegraphics[width=5.9cm]{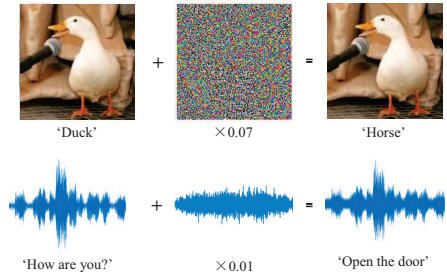}
  \caption{An illustration of machine learning adversarial examples. Studies have shown that by adding an imperceptibly small, but carefully designed perturbation, an attack can successfully lead the machine learning model to making a wrong prediction. Such attacks have been used in computer vision (upper graphs)~\cite{18szegedy2013intriguing} and speech recognition (lower graphs)~\cite{06cisse2017houdini,05gong2017crafting,12carlini2018audio}.}
  \label{fig:adv}
\end{figure}

\subsubsection{Machine Learning Level Attack}

State-of-the-art voice controlled systems are usually equipped with an automatic speech recognition (ASR) algorithm to convert digital speech signal to text. Deep neural network (DNN) based algorithms such as DeepSpeech~\cite{17hannun2014deep} can achieve excellent performance with around 95\% word recognition rate and hence dominate the field. However, recent studies show that machine learning models, especially DNN based models, are vulnerable to attacks by adversarial examples~\cite{18szegedy2013intriguing}. That is, machine learning models might mis-classify perturbed examples that are only slightly different from correctly classified examples (illustrated in Figure~\ref{fig:adv}). In speech, adversarial samples can sound like normal speech, but will actually be recognized as a completely different malicious command by the machine, e.g., an audio file might sound like ``hello'', but will be recognized as ``open the door'' by the ASR system. In recent years, several examples of such attacks have been studied~\cite{03carlini2016hidden,05gong2017crafting,12carlini2018audio,02vaidya2015cocaine,06cisse2017houdini,alzantot2018did}.

Cocaine Noodles~\cite{02vaidya2015cocaine} and Hidden Voice Command~\cite{03carlini2016hidden} are the first efforts to utilize the differences in the way humans and computers recognize speech and to successfully generate adversarial sound examples that are intelligible as a specific command to ASR systems (Google Now and CMU Sphinx), but are not easily understandable by humans. The authors observe that ASR systems rely on acoustic features (e.g., Mel-frequency cepstral coefficients or MFCC) extracted from the audio. Therefore, their attack mangles a malicious voice command signal in such a way that it retains enough acoustic features for the ASR system to accept it while making it difficult for humans to understand. The limitation of the approach in~\cite{02vaidya2015cocaine,03carlini2016hidden} is that the generated audio does not sound like legitimate speech. As a matter of fact, strictly speaking, Cocaine Noodles and Hidden Voice Commands are not really machine learning adversarial examples, because they use sounds that are not similar to legitimate speech~\cite{19carlini2017towards}. A user might notice that the malicious sound is an abnormal condition and may take counteractions. Further, generating such adversary examples requires a subjective human test step to ensure that it is imperceptible by humans, which makes the approach not fully automated. 

In contrast to~\cite{02vaidya2015cocaine,03carlini2016hidden}, more recent efforts~\cite{06cisse2017houdini,05gong2017crafting,12carlini2018audio} take advantage of an intriguing property of DNN by generating malicious audio that sounds almost completely like normal speech by adopting a mathematical optimization method. The goal of these techniques is to design a minor perturbation in the speech signal that can fool an ASR system. In~\cite{12carlini2018audio}, the authors propose a method that can produce an audio waveform that is less than 0.1\% different from a given audio waveform, but will be transcribed as any desired text by DeepSpeech~\cite{17hannun2014deep}. In~\cite{05gong2017crafting}, the authors demonstrate that a 2\% designed distortion of speech can make state-of-the-art DNN models fail to recognize the gender and identity of the speaker. In~\cite{06cisse2017houdini}, the authors show that such attacks are transferable to different and unknown ASR models. Such attacks are dangerous, because users do not expect that normal speech samples, such as ``hello'', could be translated into a malicious command by an IoT device. Despite their extraordinary performance, these methods have a clear limitation: since the perturbation is very minor, it might not be correctly captured over the air by the victim device and hence fail to attack. In~\cite{12carlini2018audio}, the authors report that their method does not have an over-the-air threat, while in~\cite{06cisse2017houdini,05gong2017crafting}, the authors do not report the over-the-air attack performance. 

\subsection{The Adversary's Knowledge}
One important factor of attacks is the adversary's knowledge. \emph{White-box attacks} assume that the adversary knows all details (e.g., design details and characteristics) of the target, while \emph{black-box attacks} assume that the adversary does not have such information. 
Table~\ref{tab:attack} summarizes several attack schemes discussed in this paper, including their type and the required adversary's knowledge.
Hardware level attacks are usually white-box attacks, because the device can be easily dis-assembled and tested, e.g., in~\cite{01zhang2017dolphinattack}, the authors first test the frequency response of the MEMS microphone and then take advantage of its nonlinearity to conduct the attack. All OS level attacks are white-box attacks. Note that all discussed schemes~\cite{08diao2014your,04alepis2017monkey,21jang2014a11y} are targeting Android devices, which is not surprising since Android is open-source and its inner workings (e.g., authority management, inter-process communication) are well known. In contrast, the OS of Amazon Echo is closed, which prevents it from being attacked. In fact, it is difficult to perform black-box hardware and OS level attacks.

Different from the OS and hardware level attacks, practical machine learning level attacks are usually black-box attacks, because state-of-the-art ASR systems for IoT devices do not release their detailed algorithms and the training sets. These ASR systems run in the cloud, where an adversary may not have access. However, machine learning attacks are able to attack unknown ASR systems, e.g., in~\cite{02vaidya2015cocaine,03carlini2016hidden,06cisse2017houdini}, the authors successfully attack Google Voice without knowing its details. This is because ASRs use similar (explicit or implicit) acoustic features and models (e.g., network architectures), which makes the machine learning adversarial examples universal. This characteristic makes machine learning level attacks more dangerous. 

Another noteworthy point is that attacks can be combined to become even more dangerous, e.g., GVS-Attacks~\cite{08diao2014your} and the approach described in~\cite{12carlini2018audio} can be combined so that the malware replay machine learning adversarial example sounds like normal speech instead of a malicious command when it finds an opportunity. Further, all attacks can be combined with the one in~\cite{04alepis2017monkey} to become an interactive attack. 

\section{Defense Strategies}

To defend and prevent OS level attacks, voice input and output need to be decoupled since simultaneously using voice input and output has been shown to affect users' security~\cite{04alepis2017monkey,08diao2014your,21jang2014a11y}. For example, AuDroid~\cite{09petracca2015audroid} has been proposed to manage the audio channel authority. By using different security levels for different audio channel usage patterns, AuDroid can resist a voice attack using the device's built-in speaker~\cite{04alepis2017monkey,08diao2014your}. However, AuDroid uses a speaker verification system to defend external replay attacks, including hardware and machine learning level attacks, which is not effective enough. Therefore, AuDroid is only robust to OS level attacks.

One defense strategy seems promising for hardware and machine learning level attacks is \emph{adversarial training}, i.e., training an extra machine learning model that can classify legitimate samples and adversaries. In~\cite{01zhang2017dolphinattack}, the authors use support vector machine (SVM) to build such a classifier that can fully defend against their proposed attack. Similarly, in~\cite{03carlini2016hidden}, the authors use logistic regression and achieve 99.8\% defense rate. A limitation of adversarial training is that it needs to know the details of the attack technology or it needs to collect a sufficient amount of adversarial examples. In practice, the attackers will not publish their approaches and they can always change the parameters (e.g., the modulation frequency in~\cite{11kasmi2015iemi} or the perturbation factor in~\cite{05gong2017crafting}) to bypass the defense. Thus, adversarial training is weak in preventing unknown attacks.

Other efforts~\cite{08diao2014your,03carlini2016hidden,09petracca2015audroid} mention the possibility of using speaker verification (SV) systems for defense. However, this is not strong enough, because the SV system itself is vulnerable to machine learning adversarial examples~\cite{05gong2017crafting} and previously recorded user speech~\cite{03carlini2016hidden,14chen2017you}.

From the perspective of the defender, a strategy that can resist various (and even unknown) attacks is expected. One observation is that all existing attacks are based on the replay attack: OS level and machine learning level attacks replay a sound and hardware level attacks replay a designed signal. In other words, the sound source is an electronic device (e.g., loudspeaker, signal generator) instead of a live speaker. On the other hand, only the command from a live speaker should be legitimate in practical applications. That is, \textbf{if we can determine if the received signal is from a live speaker, we can defend all above mentioned and even unknown attacks}. To the best of our knowledge, such techniques are non-trivial and have not been widely studied. As an approximation, in~\cite{07lei2017insecurity}, the authors propose a virtual security button (VSButton) that leverages Wi-Fi technology to detect indoor human motions and voice commands are only accepted when human motion is detected. The limitation of this work is that voice commands are not necessarily accompanied with detectable motion. In~\cite{10feng2017continuous}, the authors propose VAuth, which collects the body-surface vibration of the user via a wearable device and guarantees that the voice command is from the user. The limitation of VAuth is the need for wearable devices (i.e., earbuds, eyeglasses, and necklaces), which may be inconvenient. Finally, in~\cite{14chen2017you}, the authors determine if the source of voice commands is a loudspeaker via a magnetometer and reject such commands. The limitation of this work is that this works only up to 10cm, which is less than the usual human-device distance. Further, this approach does not work for unconventional loudspeakers and malicious signal generators. In summary, existing defense techniques are able to address only some vulnerabilities and therefore more powerful defense techniques will be required to protect voice-driven IoT devices.

%

\bibliographystyle{IEEEtran}
\bibliography{bare_conf}

\end{document}